\documentclass[a4paper,11pt]{article}
\pdfoutput=1

\usepackage[a4paper,
  left=2.5cm, right=2.5cm,
  top= 3cm, bottom=4cm]{geometry}

\usepackage{amsmath}
\numberwithin{equation}{section}
\usepackage{mathtools}
\usepackage{oldgerm}
\usepackage{amssymb}
\usepackage{bbm}
\usepackage{graphicx}
\usepackage{subcaption}
\usepackage{tikz-cd} 
\usetikzlibrary{matrix,arrows,decorations.pathmorphing}
\usepgfmodule{shapes}

\usetikzlibrary{arrows,decorations.markings}
\usepackage{color}
\usepackage{pgf}
\usepackage{pgffor}

\usepackage{epic}
\usepackage{hyperref}
\usepackage{bbold}

\newcommand{\beqa}{\begin{eqnarray}}
\newcommand{\eeqa}{\end{eqnarray}}

\def\Tr{\rm Tr}

\newcommand{\beq}{\begin{equation}}
\newcommand{\eeq}{\end{equation}}
\newcommand{\bea}{\begin{eqnarray}}
\newcommand{\eea}{\end{eqnarray}}

\newcommand{\CN}{{\mathcal N}}

\newcommand{\CT}{{\mathcal T}}

\newcommand{\CW}{{\mathcal W}}

\newcommand\qt{\tilde q}
\newcommand\bt{\tilde b}

\newcommand{\be}{\begin{equation}}
\newcommand{\ee}{\end{equation}}

\newcommand{\bpic}{\begin{tikzpicture}}
\newcommand{\epic}{\end{tikzpicture}}

\begin{document}

\thispagestyle{empty}

\begin{center}
{\Huge RG flows with supersymmetry enhancement and geometric engineering} 
\\[15mm]
{\Large{Simone Giacomelli}  
} 
\vskip 6mm
 
\bigskip
{\it  
International Center for Theoretical Physics, Strada Costiera 11, 34151 Trieste, Italy.
\\
INFN, Sezione di Trieste,
Via Valerio 2, 34127 Trieste, Italy 
\\
  }
\vskip 6 mm
e-mail: sgiacome@ictp.it
\bigskip
\bigskip

{\large{\bf Abstract}}\\[5mm]
{\parbox{14cm}{\hspace{5mm}

In this paper we study a class of $\mathcal{N}=2$ SCFTs with ADE global symmetry defined via Type IIB compactification on a class of hypersurfaces in $\mathbb{C}^3\times\mathbb{C}^*$. These can also be constructed by compactifying the 6d (2,0) theory of type ADE on a sphere with an irregular and a full punctures. When we couple to the ADE moment map a chiral multiplet in the adjoint representation and turn on a (principal) nilpotent vev for it, all the theories in this family display enhancement of supersymmetry in the infrared. We observe that all known examples of theories which flow, upon the same type of deformation, to strongly coupled $\mathcal{N}=2$ theories fit naturally in our framework, thus providing a new perspective on this topic. We propose an infrared equivalence between this RG flow and a manifestly $\mathcal{N}=2$ preserving one and, as a byproduct, we extract a precise prescription to relate the SW curves describing the UV and IR fixed points for all theories with A or D global symmetry. We also find, for a certain subclass, a simple relation between UV and IR theories at the level of chiral algebras.  
}
}
\end{center}
\newpage
\pagenumbering{arabic}
\setcounter{page}{1}
\setcounter{footnote}{0}
\renewcommand{\thefootnote}{\arabic{footnote}}

\tableofcontents

\section{Introduction} 

Four dimensional theories with eight supercharges represent an extremely important class of QFT models, since many properties can be computed exactly and they display many interesting connections with other models in diverse dimensions and with geometry. Shortly after the discovery of the Seiberg-Witten solution \cite{Seiberg:1994rs,Seiberg:1994aj} several examples of nonlagrangian theories\footnote{By this we mean that the structure of the Coulomb Branch (e.g. the presence of Coulomb Branch operators of fractional dimension) is not compatible with that of any gauge theory.} were found (see e.g. \cite{Argyres:1995jj,Argyres:1995xn,Eguchi:1996vu,Eguchi:1996ds}). In the past twenty years many other examples of nonlagrangian theories were found and more recently the class $\mathcal{S}$ construction \cite{Gaiotto:2009we} provided a general framework to study a vast landscape of nonlagrangian theories. 

Recently in \cite{Maruyoshi:2016tqk,Maruyoshi:2016aim,Agarwal:2016pjo} (see also \cite{Gadde:2015xta} for a complementary approach) it was realized that the nonlagrangian class of $\mathcal{N}=2$ theories is actually smaller than what we thought, if we relax the assumption that the gauge theory description has manifest extended supersymmetry: the authors noticed that if we consider an $\mathcal{N}=2$ lagrangian SCFT with a global symmetry $G$, couple a chiral multiplet transforming in the adjoint representation to the $G$ moment map and turn on for it a nilpotent vev, sometimes supersymmetry enhances in the IR. This construction always features the decoupling of operators which hit the unitarity bound and the resulting IR fixed point is actually a $\mathcal{N}=2$ SCFT plus decoupled chiral multiplets. 

The procedure can actually be improved to obtain a lagrangian description of the strongly-coupled SCFT alone, without any decoupled sector \cite{Benvenuti:2017lle} and this is crucial to recover supersymmetry enhancement in lower dimension by compactification of the lagrangian theory \cite{Benvenuti:2017kud,Benvenuti:2017bpg}. The construction works as follows: we start by applying the method proposed by Maruyoshi and Song and identify the set of chiral operators $\mathcal{O}_i$ which violate the unitarity bound in the IR with a run of a-maximization \cite{Intriligator:2003jj}. Then, we introduce by hand chiral multiplets $\beta_{\mathcal{O}_i}$ and turn on the superpotential terms $\mathcal{W}=\beta_{\mathcal{O}_i}\mathcal{O}_i$. The F-terms for $\beta_{\mathcal{O}_i}$ set to zero in the chiral ring the ``offending'' operators. The resulting theory is not affected by unitarity bound violations and flows directly in the IR to the model with enhanced supersymmetry. We will refer to this construction as susy enhancing RG flow or susy enhancing procedure. 

The main tool used in \cite{Maruyoshi:2016tqk,Maruyoshi:2016aim,Agarwal:2016pjo} is a-maximization, which allows to identify the R-symmetry of the infrared fixed point. For some theories and only for some choice of the nilpotent vev the R-charges in the infrared are rational and in this case the resulting fixed points can almost always be identified with known strongly-coupled $\mathcal{N}=2$ SCFTs. This procedure works only in very special cases without a recognizable pattern and at present we do not have a general criterion saying when one should expect supersymmetry enhancement (see however \cite{Evtikhiev:2017heo}, which discusses the necessary and the sufficient conditions for supersymmetry enhancement at the level of the superconformal index) and even when this happens, it is not clear how to identify the IR SCFT in advance without going through the details of the a-maximization computation. The purpose of this note is to make progress in this direction by proposing a uniform and more systematic approach to the study of susy enhancing RG flows. Our basic observation is that the geometric engineering setup is a convenient framework to formulate this question. 

Our construction goes as follows: we start by considering the set of class $\mathcal{S}$ generalized Argyres-Douglas theories defined by compactification of the 6d $\mathcal{N}=(2,0)$ theory of type $J=ADE$ on a sphere with one irregular puncture. These were classified in \cite{Wang:2015mra} and in the same paper it was observed that these models can be described in the context of geometric engineering\footnote{The methods used in this paper allow to describe in Type IIB a subset of class $\mathcal{S}$ theories on a punctured sphere. For example, in the $A_{N}$ case it is known that one can have three classes of irregular punctures \cite{Xie:2012hs}. In the present work we will discuss only theories with Type I and Type II irregular punctures (in the notation of \cite{Xie:2012hs}). It would be important to understand how to incorporate Type III punctures as well.} by compactifying Type IIB on a hypersurface singularity in $\mathbb{C}^4$. For given $J$, the theories we obtain in this way are labelled by two integers $b$ and $k$ which specify the choice of irregular puncture: $b$ can take two or three different values depending on the specific Lie algebra considered and $k$ is an arbitrary positive integer. The resulting models were dubbed $J^b(k)$ in \cite{Wang:2015mra}. 

In the same spirit of \cite{Cecotti:2012jx,Cecotti:2013lda}, we can obtain from $J^b(k)$ models a large class of theories with (at least) $J$ global symmetry by considering the same hypersurfaces in $\mathbb{C}^3\times\mathbb{C}^*$ rather than $\mathbb{C}^4$. The resulting theories are labelled by the same data ($b$ and $k$) and they also have a class $\mathcal{S}$ realization: the UV curve is again the sphere with one irregular puncture and a full regular puncture of type $J$. We call these models $D_k^b(J)$ theories since they constitute a generalization of $D_k(J)$ theories studied in \cite{Cecotti:2013lda}\footnote{I would like to mention the fact that, although these theories generalize the $D_k(J)$ class, all of them can be obtained starting from $D_k(J)$ theories and turning on a relevant $\CN=2$ preserving deformation.}. Since these theories have a nonabelian global symmetry, we can consider the susy enhancing RG flow for them. 

In this note we conjecture that for all the models in this class supersymmetry always enhances in the infrared upon turning on a principal nilpotent vev for the $J$ global symmetry and the IR fixed point of the resulting RG flow turns out to have a very simple geometric description in Type IIB: it is the $J^b(k)$ theory described by the same quasi-homogeneous equation as the parent UV theory, but in $\mathbb{C}^4$ rather than $\mathbb{C}^3\times\mathbb{C}^*$. Overall, the net effect of the RG flow is simply to change the ambient space by introducing the $\mathbb{C}^3$ hyperplane at the origin. Remarkably, this construction allows to recover all known RG flows with supersymmetry enhancement and in particular captures all examples of ``lagrangians for nonlagrangian theories'': these are simply recovered by focusing on the subclass of lagrangian $D_k^b(J)$ theories. Furthermore, our geometric setup provides infinitely many new nonlagrangian examples, as we will explain later. We provide evidence for our conjecture by observing that the a-maximization analysis is perfectly consistent with the above claim.

Another interesting feature of this approach is to remove the ambiguity in the choice of nilpotent vev which is part of the data defining the procedure developped in \cite{Maruyoshi:2016tqk,Maruyoshi:2016aim,Agarwal:2016pjo}: for a given theory only some special choices of nilpotent vev induce supersymmetry enhancement and at present there is no known characterization of these ``distinguished'' nilpotent orbits. This information is automatically captured by our geometric setup, in the sense that in general a single $\mathcal{N}=2$ SCFT has multiple realizations in the $D_k^b(J)$ class and the manifest global symmetry $J$ is only a subgroup of the full symmetry group of the theory. Therefore, a principal nilpotent vev for $J$ corresponds in general to a non principal nilpotent vev for the full global symmetry of the theory. We find that uniformly considering only the $J$ principal nilpotent vev for $D_k^b(J)$ theories we nevertheless recover all the RG flows with supersymmetry enhancement identified in \cite{Agarwal:2016pjo}. In other words, in our geometric setup we do not miss RG flows displaying supersymmetry enhancement by focusing on principal nilpotent vevs only.

Further insight on these RG flows is provided by combining our observation with the class $\mathcal{S}$ realization of the models considered in the present paper: starting from a $D_k^b(J)$ model the IR fixed points of the susy enhancing RG flow can also be realized by closing completely the full puncture (i.e. giving a nilpotent vev to the moment map)  of another $D_k^b(J)$ theory (with different $b$) and consequently we find an infrared duality between the susy enhancing RG flow and a manifestly $\mathcal{N}=2$ preserving RG flow (in this case a higgsing). This result has interesting consequences, for example it allows to extract a simple relation between the Seiberg-Witten data (meaning curve and differential) of the UV and IR theories. Another interesting outcome is a relation at the level of the corresponding 2d chiral algebras \cite{Beem:2013sza}, at least for $D_k^b(J)$ models whose global symmetry is exactly $J$ with no further enhancement. All these relations between UV and IR theories are not manifest in other approaches, since the RG flow breaks extended supersymmetry at intermediate scales and the above quantities are defined only for $\mathcal{N}=2$ theories. In this sense our duality is instrumental in deriving them.

The paper is organized as follows: in sections 2 we review the geometric engineering setup for generalized Argyres-Douglas theories and discuss the properties of $D_k^b(J)$ theories. We compute several quantities which are needed in later sections and in section 2.3 we formulate precisely our duality statement, the key observation of the present work, and discuss general properties of the susy enhancing RG flows. These are derived from the a-maximization analysis which is by now standard and we review it in detail in the Appendix. Section 3 is devoted to the discussion of $D_k^b(J)$ theories for low values of $k$ and also those with $J=SU(2)$. We identify their Seiberg-Witten (SW) curves for $J=A,D$ and then we proceed with the analysis of the lagrangian subclass, which coincides precisely with all the quiver theories discussed in \cite{Maruyoshi:2016tqk,Maruyoshi:2016aim,Agarwal:2016pjo,Benvenuti:2017bpg,Agarwal:2017roi}. The main outcomes of our construction and the applications of our duality are discussed in Section 4, which constitutes the main part of the present paper: we first check that all the RG flows with supersymmetry enhancement found so far in the literature fit in our framework and provide several other examples of susy enhancing RG flows. We explain how to relate SW curves of UV and IR fixed points of the susy enhancing RG flow and comment on their relation at the level of chiral algebras. Finally, in section 5 we summarize our findings and discuss possible future directions of investigation. The counting of mass parameters for $D_k^b(J)$ theories is presented in the Appendix.

\section{$\CN=2$ SCFT's from Type IIB and statement of the result}\label{geomsec} 

In this section we discuss class $\mathcal{S}$ SCFTs obtained by compactifying the $\CN=(2,0)$ theory on a sphere with one irregular puncture or one irregular puncture and a full puncture and their Type IIB realization. Since several quantities of simply-laced algebras enter in our analysis, we report them here for convenience of the reader: 

\be\begin{array}{|c|c|c|c|c|}
\hline 
\text{Algebra} & \text{Rank} & \text{Coxeter Number} & \text{Dimension} & \text{Degree of Casimir invariants} \\
\hline 
A_{N-1} & N-1 & N & N^2-1 & 2,3,\dots,N \\
\hline 
D_N & N & 2N-2 & N(2N-1) & 2,4,\dots,2N-2;N \\
\hline 
E_6 & 6 & 12 & 78 & 2,5,6,8,9,12 \\
\hline
E_7 & 7 & 18 & 133 & 2,6,8,10,12,14,18 \\
\hline 
E_8 & 8 & 30 & 248 & 2,8,12,14,18,20,24,30 \\
\hline
\end{array}\ee

\subsection{Class $\mathcal{S}$ theories on the sphere with one irregular puncture and geometric engineering}

As we have mentioned in the introduction, by compactifying the $\CN=(2,0)$ six dimensional theory of type $J$ on a sphere with one irregular puncture one can get a large class of $\CN=2$ superconformal theories in four dimensions, which were classified in \cite{Wang:2015mra}. Following the notation of \cite{Xie:2016evu}, we call the resulting models $J^b(k)$, where $b$ and $k$ are integers parametrizing the choice of irregular puncture. These theories can also be realized in the context of geometric engineering by compactifying Type IIB string theory on a three-fold hypersurface singularity in $\mathbb{C}^4$ $W(x_1,x_2,x_3,z)=0$. The set of relevant geometries is given in \cite{Wang:2015mra} and we list them here for convenience: 
\be\label{sing}\begin{array}{|c|c|c|}
\hline
J & \text{Singularity} & b \\
\hline 
A_{N-1} & x_1^2+x_2^2+x_3^N+z^k=0& N \\
\hline
& x_1^2+x_2^2+x_3^N+x_3z^k=0& N-1 \\
\hline 
D_N & x_1^2+x_2^{N-1}+x_2x_3^2+z^k=0& 2N-2 \\
\hline 
 & x_1^2+x_2^{N-1}+x_2x_3^2+x_3z^k=0& N \\
\hline 
E_6 & x_1^2+x_2^3+x_3^4+z^k=0& 12 \\
\hline
 & x_1^2+x_2^3+x_3^4+x_3z^k=0& 9 \\
\hline
 & x_1^2+x_2^3+x_3^4+x_2z^k=0& 8 \\ 
\hline 
E_7 & x_1^2+x_2^3+x_2x_3^3+z^k=0& 18 \\
\hline
 & x_1^2+x_2^3+x_2x_3^3+x_3z^k=0& 14 \\ 
\hline 
E_8 & x_1^2+x_2^3+x_3^5+z^k=0& 30 \\
\hline 
 & x_1^2+x_2^3+x_3^5+x_3z^k=0& 24 \\
\hline 
 & x_1^2+x_2^3+x_3^5+x_2z^k=0& 20 \\
\hline
\end{array}\ee

Notice that when $b=h(J)$ (the (dual) Coxeter number of $J$) the singularities listed above are precisely those defining the $(A_{k-1},J)$ theories discussed in \cite{Cecotti:2010fi}. One further piece of information we need is the holomorphic three-form, which for the present class of theories can be written in the form 
\be\label{cyform}\Omega=\frac{dz\prod_idx_i}{dW}.\ee 
Since the integral of $\Omega$ measures the mass of BPS particles, we should require $\Omega$ to have dimension one and this fact can be used to extract the scaling dimension of the various coordinates and hence the dimension of Coulomb branch (CB) operators as well, which appear as complex deformation parameters in the geometric engineering setup. One way to identify them is to consider the polynomial ring generated by the variables $x_i$ and $z$ modulo the ideal $\mathcal{I}_W$ generated by the polynomials $\frac{\partial W}{\partial x_i}$ and $\frac{\partial W}{\partial z}$. The set of allowed complex deformations is given by the quotient algebra \cite{Shapere:1999xr} 
\be\label{quotient}\mathcal{A}_W=\mathbb{C}[x_1,x_2,x_3,z]/\mathcal{I}_W.\ee
From the geometric engineering setup one can also extract the $c$ central charge of $J^b(k)$ theories \cite{Xie:2016evu}: 
\be\label{centralc}\begin{array}{|c|c|c|}
\hline 
A_{N-1} & \frac{(N-1)(k-1)(N+k+Nk)}{12N+12k}\;(b=N) & \frac{(Nk-N+1)(N+k+Nk-1)}{12(N+k-1)}\;(b=N-1) \\
\hline 
D_N & \frac{N(k-1)(2kN+2N-k-2)}{12(2N-2+k)}\;(b=2N-2) & \frac{(2Nk-N-2k)(2Nk+N-k)}{12N+12k}\;(b=N)  \\
\hline 
E_6 & \frac{(k-1)(13k+12)}{2k+24}\;(b=12) & \frac{(4k-3)(13k+9)}{6k+54}\;(b=9) \\
\hline 
 & \frac{(3k-2)(13k+8)}{4k+32}\;(b=8) & \\
\hline
E_7 & \frac{7(k-1)(19k+18)}{12(k+18)}\;(b=18) & \frac{(9k-7)(19k+14)}{12(k+14)}\;(b=14) \\
\hline 
E_8 & \frac{2(k-1)(31k+30)}{3k+90}\;(b=30) & \frac{(5k-4)(31k+24)}{6(k+24)}\;(b=24) \\ 
\hline
 & \frac{(3k-2)(31k+20)}{3(k+20)}\;(b=20) & \\
\hline
\end{array}\ee 
Actually, the numbers appearing in the above table coincide with the $c$ central charge of $J^b(k)$ theories only when the theory has no global symmetries (and accordingly no mass parameters, which correspond to deformation parameters of dimension one). We point out that whenever this constraint is not satisfied, formula (\ref{centralc}) should be ``adjusted'' by subtracting $\frac{G_F}{12}$, where $G_F$ is the number of mass parameters (or equivalently the rank of the global symmetry of the theory). The counting of mass parameters is performed in the Appendix. 

In order to understand the origin of the last statement, we need to go back to the derivation of (\ref{centralc}) proposed in \cite{Xie:2015rpa}. The analysis builds on the result of \cite{Shapere:2008zf} 
\be\label{ctopo} c=\frac{R(B)}{3}+\frac{r}{6}+\frac{h}{12},\ee 
where $h$ is the number of free hypermultiplets (in our case $h=0$)\footnote{This was argued in section 2.4 of \cite{Xie:2015rpa} exploiting the fact that at generic points of the Coulomb Branch (i.e. in the deformed singularity) the only nontrivial homology groups in the geometry are $H^0$ and $H^3$, therefore only vector multiplets (arising from the RR four-form $C_4$) and no massless hypermultiplets are generated in the compactification of Type IIB.}, $r$ is the dimension of the Coulomb Branch and $R(B)$ is the scaling dimension of the discriminant of the Seiberg-Witten (SW) curve divided by four. It was argued in \cite{Xie:2015rpa} that for this class of models \be\label{conjc} R(B)=\frac{1}{4}\mu D(u_{max}),\ee 
where $\mu=2r+G_F$ (the Milnor number) is the rank of the charge lattice (or equivalently the number of nodes of the underlying BPS quiver \cite{Cecotti:2010fi}) and $D(u_{max})$ is the scaling dimension of the CB operator with largest dimension. 
It can be shown that for $J^b(k)$ theories $$D(u_{max})=\frac{kh(J)}{k+b},$$ 
so (\ref{ctopo}) becomes 
\be c=\frac{\mu}{12}\left(\frac{kh(J)}{k+b}+1\right)-\frac{G_F}{12}.\ee 
Plugging in the formula
\be\label{milnum1}\mu=\frac{r(J)}{b}(kh(J)-b),\ee
one finds precisely the values appearing in (\ref{centralc}) minus $\frac{G_F}{12}$.

Once we know $c$, the central charge $a$ can be extracted from the scaling dimension of CB operators exploiting the relation \cite{Shapere:2008zf} \be\label{centrala} 8a-4c=\sum_i(2D(u_i)-1).\ee

\subsection{SCFT's with $A,D,E$ global symmetry from Type IIB string theory}\label{scftj}

Starting from the above list of three-fold singularities (or $\CN=2$ SCFT's), we can construct another class of SCFT's labelled by the same data, such that the Lie group $J$ actually corresponds to (in general a subgroup of) the global symmetry of the theory. These can be defined by compactifying Type IIB on the singularities listed above with the modification $$z\rightarrow e^z,$$ 
so that we are now dealing with a hypersurface singularity in $\mathbb{C}^3\times\mathbb{C}^*$. The holomorphic three-form is always given by (\ref{cyform}). By working in terms of the $\mathbb{C}^*$ variable $t=e^z$, we can bring the geometry back to the form (\ref{sing}) (with $z$ replaced by $t$) but indeed the theories are not the same. One quick way to see this is to notice that the Calabi-Yau structure is different: in the coordinates $(t,x_i)$ the holomorphic three-form reads 
$$\Omega=\frac{dt\prod_idx_i}{tdW}$$
and we clearly see that the assignment of scaling dimensions for the various coordinates is now completely different. In particular, requiring $\Omega$ to have dimension one results in an assignment of scaling dimension for the coordinates $x_i$ which does not depend on $k$ and $b$. The result is reported in the following table: 
\be\label{scaledim}\begin{tabular}{|c|c|c|c|}
\hline 
 & $x_1$ & $x_2$ & $x_3$ \\
\hline 
$A_{N-1}$ & $\frac{N}{2}$ & $\frac{N}{2}$ & 1 \\
\hline
$D_N$ & $N-1$ & 2 & $N-2$ \\
\hline 
$E_6$ & 6 & 4 & 3 \\
\hline 
$E_7$ & 9 & 6 & 4 \\
\hline 
$E_8$ & 15 & 10 & 6 \\ 
\hline
\end{tabular}\ee
We can also easily derive the dimension of the coordinate $t$ and the singularity $W$: 
\be\label{scalet} D(t)=\frac{b}{k};\quad D(W)=h(J).\ee 
The complex structure deformations corresponding to CB operators can be identified from the quotient algebra (\ref{quotient}). There are two differences with respect to the $J^b(k)$ class one should take into account: 
\begin{enumerate}
\item the ideal $I_W$ is now generated by $\frac{\partial W}{\partial x_i}$ and $t\frac{\partial W}{\partial t}$ (so for e.g. $J=SU(N)$ and $b=N$ terms proportional to $t^{k-1}$ are now allowed).
\item Versal deformations of the ADE singularity (which do not depend on $t$) are interpreted as the Casimir invariants of the $J$ mass matrix. These do not correspond to CB operators but are rather interpreted as mass parameters.
\end{enumerate}
Taking this into account we clearly see that the CB operator of maximal dimension (the corresponding term is $\delta W=ut$) has dimension \be\label{dimmax} D(u)=h(J)-\frac{b}{k}.\ee
We will call these theories $D_k^b(J)$ and when $b=h(J)$ these coincide with $D_k(J)$ theories studied in \cite{Cecotti:2012jx},\cite{Cecotti:2013lda} so we will refer to them with this name below. 

Let's now pause to notice that the scaling dimensions of the coordinates $x_i,t$ for $D_k^b(J)$ theories and $x_i,z$ for $J^b(k)$ theories are closely related: the dimension of $x_i,z$ can be obtained from (\ref{scaledim}) and (\ref{scalet}) simply by multiplying everything by $\frac{k}{k+b}$. It is easy to see that with this assignment the three-form $\Omega$ in (\ref{cyform}) has dimension one. This observation will be useful below.

For $D_k^b(J)$ theories $\mu\equiv2r+G_F$ is equal to 
\be\label{milnum2}\mu=k\frac{r(J)h(J)}{b},\ee 
where $r(J)$ denotes the rank of the group $J$. Notice that $b$ (see Table (\ref{sing})) is always a divisor of $r(J)h(J)$. Since the theory always has at least global symmetry $J$, it is convenient to rewrite the rank of the global symmetry $G_F$ as $r(J)+n$, where $n$ is the number of the remaining mass parameters if any. The value of $n$ is computed in the Appendix.
Using now (\ref{ctopo}) and (\ref{conjc}), we conclude that the $c$ central charge of $D_k^b(J)$ theories is 
\be\label{centralc2} c=\frac{r(J)}{12b}(kh(J)-b)(h(J)+1)-\frac{n}{12}.\ee
Setting $n=0$ and replacing $k$ with $k+b$ (we will comment below on the reasons underlying this shift), (\ref{centralc2}) reproduces the value for the $c$ central charge of $(J^b(k),F)$ theories tabulated in \cite{Xie:2016evu}. Once we know $c$, the $a$ central charge can be computed using (\ref{centrala}).

There is a simple field theoretic connection between $J^b(k)$ and $D_k^b(J)$ which will be rather important in the present paper: as we explained before, $D_k^b(J)$ theories have global symmetry $J$, so there is a corresponding moment map and by giving a (principal) nilpotent vev to it\footnote{This statement refers to the case $k>b$.} we break spontaneously the $J$ global symmetry. By turning on this vev we initiate an RG flow which lands (for $k>b$) precisely on the $J^b(k)$ class. More explicitly, the UV and IR fixed points of the above mentioned RG flow are 
$$D_{k+b}^b(J)\rightarrow J^b(k)\quad \forall k>0.$$ 
This fact has a natural class $\mathcal{S}$ interpretation if we notice that all $D_k^b(J)$ theories with $k>b$ can also be realized by compactifying the 6d $\CN=(2,0)$ theory of type $J$ on a sphere with one irregular puncture (the same irregular puncture which engineers the model $J^b(k-b)$) and a regular full puncture. Equivalently, we have for every $k>0$ the relation $D_{k+b}^b(J)=(J^b(k),F)$ in the notation of \cite{Xie:2016evu} (this explains the shift in $k$ mentioned before). From this perspective the RG flow described above just corresponds to removing the full puncture\footnote{This operation is usually called closure of the puncture in the literature}. This description is allowed but less suited for the case $k\leq b$, since closing the full puncture in this case involves subtleties. We would like to remark that the models we get for $k$ small often have a class $\mathcal{S}$ realization based on a 6d theory of lower rank. The lagrangian subclass we will discuss in the next section is a clear example. 

In \cite{Xie:2016evu} some properties of $(J^b(k),F)$ theories were discussed. In particular, it was conjectured that the flavor central charge of the $J$ global symmetry is equal to the $U(1)_R$ charge of the CB operator of largest dimension. For all $D_k^b(J)$ theories this is equal to (see \ref{dimmax})
$$2h(J)-2\frac{b}{k},$$ 
which is precisely the answer proposed in \cite{Xie:2016evu}. A simple trick to guess this value for the flavor central charge is the following: the Calabi-Yau geometry corresponding to a $J$ vectormultiplet coupled to $D_k^b(J)$ can be uniformely written in the following form 
\be\label{gaugeddp} W(x_i,t)+\frac{\Lambda^{b_1}}{t}=0,\ee 
where $\Lambda$ is a constant which is physically interpreted as the dynamically-generated scale of the gauge theory and $b_1$ is the one loop coefficient of the beta function. This is in turn equal to $2h(J)-\beta$ where $\beta$ is the contribution to the beta function from the $D_k^b(J)$ theory (and indeed this is half the flavor central charge). By formally imposing that all the terms have the same ``dimension'' and assigning dimension one to $\Lambda$ we find the value of $b_1$\footnote{A more rigorous derivation, from which our ``heuristic argument'' is derived, is obtained adapting the computation performed in Section 2 of \cite{Tachikawa:2011yr}: our CY geometry is given by an ADE singularity fibered over the cylinder parametrized by $z$. We can now introduce a scale $a$ and deform the ADE singularity imposing on the corresponding Casimirs $w_i$ the relation $w_i\sim a^{d_i}$ ($d_i$ denotes the degree of the Casimir). The mass of the ADE W-bosons is obtained integrating $\Omega$ on the two-cycles of the ADE singularity $C_i$ and over a cycle wrapping once around the cylinder. The mass of a monopole is given instead by the integral of $\Omega$ over $C_i$ and a ``radial'' cycle on the cylinder interpolating between the regions at small and large $\vert z\vert$ where the cycles $C_i$ shrink. The ratio of these masses gives the gauge coupling, which in turn gives the beta function of the ADE gauge theory when we differentiate with respect to $a$. Following this procedure we recover (\ref{betacon}).}: from (\ref{scalet}) we immediately conclude \be\label{betacon}b_1=h(J)+\frac{b}{k}\rightarrow \beta=h(J)-\frac{b}{k}.\ee 
For $D_k(J)$ theories the above formula reduces to 
$$h(J)\frac{k-1}{k}$$ 
which is the correct value found in \cite{Cecotti:2012jx}. We will not attempt to derive (\ref{betacon}) from the 4d/2d correspondence of \cite{Cecotti:2010fi} as was done in detail in \cite{Cecotti:2013lda} for $D_k(J)$ theories, although it should be possible given the Type IIB origin of our models. It would be important to fill in this gap.

\subsection{An infrared duality for the susy enhancing RG flow}\label{dualsec}

We now have all the ingredients we need to state the main claim of this note: starting from any $D_k^b(J)$ theory (for every choice of $J$, $b$ and $k>0$), the susy enhancing RG flow triggered by a principal nilpotent vev for the symmetry group $J$ leads to supersymmetry enhancement in the infrared and the IR fixed point is the $J_k(b)$ theory, which can also be obtained, as explained above, by closing the $J$ full puncture of $D_{k+b}^b(J)$. We are thus proposing an infrared duality: in one duality frame the RG flow is manifestly $\CN=2$ preserving (the closure of a puncture is actually a Higgs branch flow) whereas on the other side supersymmetry enhances only at long distances\footnote{Notice that in both cases we are breaking the global symmetry $J$ spontaneously, which implies the presence of Goldstone multiplets (equal in number on both sides of the duality) besides the interacting $\CN=2$ theory.}. 

We have a slightly different perspective on the above duality exploiting the fact that there is an $\mathcal{N}=2$ preserving RG flow from $D_{k+b}^b(J)$ to $D_k^b(J)$. Geometrically this can be described by a suitable deformation of the $D_{k+b}^b(J)$ geometry. For $k<b^2$ this is interpreted as giving an expectation value to a CB operator (therefore we are moving on the Coulomb branch) and for $k>b^2$ it is interpreted as a relevant deformation of the prepotential. As a result, if we first perform this deformation and then activate the susy enhancing RG flow, we flow in the IR to the same theory we land on by closing the $J$ full puncture. We thus actually get two different descriptions of the same RG flow; one is manifestly $\mathcal{N}=2$ preserving whereas the other involves two steps.

In the rest of the paper we will give evidence for this claim. We would like to mention that a special case of our duality (basically the above statement for the $D_k(SU(N))$ class) was already noticed in \cite{Maruyoshi:2016aim}: more precisely, the authors observed that $(I_{N,k},F)=D_{k+N}SU(N)$ theories flow to $I_{N,k+N}=(A_{N-1},A_{N+k-1})$ theories in the IR when applying the Maruyoshi-Song procedure. In the present work we add the observation that $I_{N,k+N}$ theories can also be obtained by closing the full puncture of $(I_{N,k+N},F)=D_{k+2N}(SU(N))$. 

\noindent We can represent pictorially our duality statement with the following diagram:
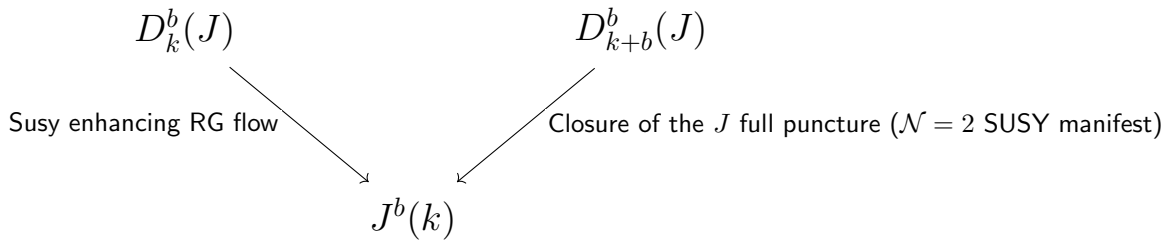
\begin{center}
\begin{tikzpicture}[node distance = 2cm, auto, inner sep=2mm]
\node  (M) at (0,0) {\Large{$J^b(k)$}};
\node  (IIB) at (-3,2.5)  {\Large{$D_k^b(J)$}};
\node  (IIA) at (3,2.5) {\Large{$D_{k+b}^b(J)$}};
 \path[every node/.style={font=\sffamily\small,
  		fill=white,inner sep=1pt}]

(IIA) edge[->] node[midway, right=7pt]{Closure of the $J$ full puncture ($\mathcal{N}=2$ SUSY manifest)} (M)
(IIB) edge[->] node[midway, left=6.5pt]{Susy enhancing RG flow} (M);

\end{tikzpicture}
\end{center}

\noindent This statement implies that the susy enhancing RG flow has a surprisingly simple description in the context of geometric engineering: as we have seen the UV theory is obtained compactifying Type IIB string theory on the CY threefold defined by the equation $W(x_1,x_2,x_3,t)=0$ (with $W$ as in (\ref{sing})) in $\mathbb{C}^3\times\mathbb{C}^*$ whereas the IR fixed point is described by the same equation in $\mathbb{C}^4$. Therefore, the net effect of the RG flow is simply to turn the ambient space into $\mathbb{C}^4$, changing the normalization of the holomorphic three-form as follows: 
\be\label{normal}\Omega_{UV}=\frac{\prod_idx_idt}{tdW}\longrightarrow \Omega_{IR}=\frac{\prod_idx_idt}{dW}.\ee 
This fact will be exploited later to relate the SW curves of the UV and IR theories. 

In the next section we will find that some of the $D_k^b(J)$ theories are lagrangian and our claim includes, as a special case, all known examples of ``lagrangian UV completions'' of AD-like theories. Moreover, our infrared duality provides a natural relation between the SW curves of the UV and IR fixed points of the susy enhancing RG flow and allows to relate the corresponding chiral algebras. 

We would now like to make the following remark: as we will see in the next section, all $D_k^b(J)$ theories with $k>b$ are not lagrangian\footnote{In this paragraph by nonlagrangian we mean that there is no ``conventional'' $\CN=2$ lagrangian description, i.e. there is no point on the conformal manifold in which the matter content is given just by free hypermultiplets and vector multiplets.}. In particular, $D_{k+b}^b(J)$ is not lagrangian for any $k$, regardless of whether $D_{k}^b(J)$ is lagrangian or not. We therefore conclude that our duality establishes the infrared equivalence of the susy enhancing RG flow for a lagrangian theory with the higgsing (closure of the puncture) of a nonlagrangian model. 

The above duality can be checked to be perfectly consistent with the a-maximization analysis, which we review in the Appendix. The most important equation is (\ref{epsmax}), which relates scaling dimensions at the UV and IR fixed points of the susy enhancing RG flow. We can immediately derive from it some general features of the susy enhancing RG flow: 
\begin{itemize}
\item For $k>b$ none of the singlets $M_i$ violate the unitarity and they all become CB operators of the IR fixed point SCFT. Conversely, for $k\leq b$ (hence for all lagrangian examples) there is at least one singlet (the one with R-charge $2+2\epsilon$) which violates the unitarity bound. 
\item We find that the number of operators which violate the unitarity bound and should be flipped is always equal to $r(J)$. Combining this with the fact that the ``candidate'' CB operators of the IR SCFT are either CB operators of the $D_k^b(J)$ theory or the singlets $M_i$ (and we always have $r(J)$ of them), we conclude that the UV and IR SCFT's always have the same rank. This conclusion also follows by comparing the Milnor numbers of the two theories (\ref{milnum1}),(\ref{milnum2}) and exploiting the fact that the rank of the global symmetry group of $D_k^b(J)$ is equal to that of $J^b(k)$ plus $r(J)$ (see the Appendix).
\end{itemize} 
The fact that the susy enhancing RG flow preserves the rank of the SCFT has not been pointed out in the literature and field-theoretically is far from being obvious. We do not have an a priori derivation of this statement however, we would like to observe that if a proof can be found, this would nicely explain why there are no known examples of supersymmetry enhancement when the UV theory has non simply-laced global symmetry. 

To illustrate our point, let us consider the case of a principal nilpotent vev for the $USp(2N-4)$ global symmetry of $SO(N)$ SQCD (these models where discussed in \cite{Agarwal:2016pjo}, with the conclusion that supersymmetry does not enhance): in this case there are $r(SO(N))-1$ singlets whose dimension is higher than that of all the $r(SO(N))$ Coulomb branch operators plus others with lower dimension. In particular, there is always a singlet which is degenerate with the CB operator of highest dimension (for example in the case $N=6$ the CB operators in the UV have dimension {2,3,4} and the singlets {2,4,6,8}). As a result, if all the CB operators decouple along the RG flow, we are left in the IR with at most $r(SO(N))-1$ singlets (which is of course strictly less than the rank of the UV theory), otherwise we end up with at least $r(SO(N))+1$ CB operators in the IR. In any case we conclude a priori that the rank of the theory cannot be preserved along the RG flow. In conclusion, proving that the rank has to be preserved whenever supersymmetry enhances would explain why these models do not work. More general nilpotent vevs and linear quivers ending with a $SO$ gauge group coupled to fundamentals can be analyzed in the same way with identical conclusions.

\section{Seiberg-Witten curves and the lagrangian subclass}

\subsection{Extracting the SW curve}

In this section we will discuss the SW curves of $D_k^b(J)$ theories\footnote{The SW curves of $J^b(k)$ theories are discussed in \cite{Wang:2015mra}.}. This is a special case of the well-known problem of extracting the Seiberg-Witten curve from the ``Seiberg-Witten geometry'' (meaning the Calabi-Yau space on which we compactify the Type IIB theory). Answering this question can be hard (see for example \cite{Lerche:1996an} which discusses this problem for $E_6$ SYM theory) and at present we do not have a general satisfactory answer. However, using techniques already available in the literature, we can extract rather easily the Seiberg-Witten curves for $D_k^b(J)$ theories at least for classical Lie groups ($J=A_{N-1}$ or $J=D_N$). This was already done in \cite{Cecotti:2013lda} for the case $b=h(J)$. The exceptional case can be handled using the method proposed in \cite{Tachikawa:2011yr} (which elaborates on the results of \cite{Lerche:1996an}), however we find it simpler to use directly the CY geometry to study this case.

The case $J=A_{N-1}$ is the simplest since the variables $x_{1,2}$ enter quadratically in (\ref{sing}) and it suffices to drop them to extract the curve\footnote{This statement is best understood in the context of the 4d/2d correspondence proposed in \cite{Cecotti:2010fi}, which relates the SCFT's engineered by the Calabi-Yau singularities (\ref{sing}) to $\CN=(2,2)$ LG models. The variables $x_i$ and $z$ are interpreted as 2d chiral multiplets and the singularity is the superpotential of the theory. In this context quadratic terms make the corresponding chiral multiplets massive hence they can be integrated out.}. 
As a result we find for $b=N$ (setting $t=e^z$)
\be x^N+t^k=0;\quad \lambda_{SW}=x\frac{dt}{t},\ee 
and for $b=N-1$ 
\be x^N+xt^k=0;\quad \lambda_{SW}=x\frac{dt}{t}.\ee 
Notice that these curves are the same as those describing $A_{N-1}^b(k)$ theories. The difference between the two classes is the normalization of the SW differential, which for $A_{N-1}^b(k)$ theories is simply $xdt$ (this observation will play an important role in the next section). The curves for the deformed theories can be easily extracted by turning on in the CY geometry all the deformations in the quotient algebra (\ref{quotient}). In this case this results in turning on all subleading terms of the form $u_{ij}x^it^j$. Notice that terms with $i=N-1$ can be removed by shifting $x$, those with $j=0$ describe the $SU(N)$ mass Casimirs and for $b=N-1$ also the term $u_{0k}t^k$ represents a mass parameter. 

The case $J=D_N$ is slightly more complicated, since now only $x_1$ appears quadratically and can be integrated out (in the same sense as in the $SU(N)$ case discussed before). This however can be circumvented, as explained in \cite{Brandhuber:1995zp}, by introducing an auxiliary variable $\lambda$ and perturbing the singularity (\ref{sing}) with the term $x_3\lambda$. Now $x_3$ is massive and can be integrated out. 

In the case of $D_k(SO(2N))$ theories this procedure leads to 
\be x_1^2+x_2^{N-1}+x_2x_3^2+t^k-x_3\lambda\longrightarrow x_2^{N-1}-\frac{\lambda^2}{x_2}+t^k.\ee 
If we now multiply everything by $x_2$ and consider the change of variable $x_2=x^2$ we get the final expression for the curve 
\be\label{dpson} x^{2N}-\lambda^2+x^2t^k=0;\quad \lambda_{SW}=x\frac{dt}{t},\ee 
where we included also the SW differential. Let us now comment about the physical interpretation of the parameter $\lambda$. As we have already explained, the $D_k(SO(2N))$ theories have $SO(2N)$ global symmetry so we can turn on the corresponding mass terms. This operation results in a deformation of the SW curve and the corresponding parameters represent the Casimir invariants of the $SO(2N)$ mass matrix. The distinctive feature of mass parameters is the fact that they appear as residues for the SW differential and this is precisely what happens when we turn on $\lambda$: the SW differential in (\ref{dpson}) has a simple pole at $t=0$ and the residue is proportional to $\lambda^{1/N}$, hence we can interpret $\lambda$ as the degree $N$ Casimir of $SO(2N)$. We then conclude that the curve describing the undeformed SCFT is  
\be x^{2N}+x^2t^k=0;\quad \lambda_{SW}=x\frac{dt}{t}.\ee 
This formula also appears in \cite{Cecotti:2013lda}. Again, by going through the above steps keeping all the complex structure deformations turned on, we can extract the deformed curve which reads 
\be\label{quiverso}x^{2N}+x^2t^k+x^2\sum_{i,j}u_{ij}x^{2i}t^j+P(t)^2=0.\ee 
In this formula $i\geq0$ and indeed $j<k$. $P(t)$ is a polynomial in $t$ of degree (at least)\footnote{For $k>N-1$ the degree of $P(t)$ may be larger than $k/2$. The corresponding parameters have dimension smaller than one and are interpreted as coupling constants associated with relevant deformations. Turning them on results in a deformation of the prepotential.} $\lfloor k/2\rfloor$ ($\lfloor..\rfloor$ denotes the integer part). The $N$ Casimirs of $SO(2N)$ are identified with the $N-1$ parameters $u_{i0}$ and the constant term in $P(t)$. For $k$ even also the term of degree $k/2$ in $P(t)$ describes a mass deformation.

Let us now repeat the above procedure for the $D_k^N(SO(2N))$ class. Adding again $\lambda$ and perturbing the singularity we find 
$$x_1^2+x_2^{N-1}+x_2x_3^2+x_3t^k-x_3\lambda.$$ 
Going through the same steps as before we conclude that the SW curve describing the undeformed theory is 
\be\label{dpsomod}x^{2N}+t^{2k}=0;\quad \lambda_{SW}=x\frac{dt}{t}.\ee 
Again we have set $\lambda=0$ since keeping it finite introduces a pole for the SW differential. The deformed curve reads 
\be\label{quiverso2}x^{2N}+x^2\sum_{i,j}u_{ij}x^{2i}t^j+(t^k+P_{k-1}(t))^2=0.\ee 
where $i\geq0$ and $j<2k$. As in the previous case, $u_{i0}$ parameters and the constant term in $P_{k-1}(t)$ are identified with the flavor Casimirs.

\subsection{Case study and the lagrangian class}\label{lagrth}

Let us now discuss the $D_k^b(J)$ models for small values of $k$ to get an intuition about what kind of theories we get and then we will proceed with the analysis of the lagrangian subclass. We will concentrate on the cases $b\neq h(J)$ since $D_k(J)$ theories have already been discussed in detail in \cite{Cecotti:2013lda}. 

In \cite{Cecotti:2013lda} it was noticed that $D_1(J)$ theories are trivial for every choice of $J$. Let us now consider the other cases with $k=1$. The SW curve of $D_1^{N-1}(SU(N))$ theories with the $SU(N)$ symmetry gauged is 
$$x^N+xt+\frac{1}{t}=0;\quad \lambda_{SW}=x\frac{dt}{t}.$$ 
This curve is well known to describe $SU(N)$ SQCD with one flavor, so we conclude that $D_1^{N-1}(SU(N))$ just describes $N$ free hypermultiplets. For $k=2$ the SW curve is $x^{N}+xt^2=0$ and if we multiply everything by $x$ and trade the variable t for $t'=tx$, the curve becomes $x^{N+1}+t'^2=0$ and the SW differential retains the canonical form $xdt'/t'$ up to exact terms. These are the SW data of $D_2(SU(N+1))$.  

The theories $D_1^{N}(SO(2N))$ are nontrivial and interacting: starting from the curve $x^{2N}+t^2=0$, with the redefinition $t=t'x^2$ and dividing everything by $x^2$ we get the SW curve of the $D_2(SO(2N-2))$ theory therefore we identify the two families. In conclusion, for $N$ even the theory is $USp(N-2)$ SQCD with $N$ fundamental hypermultiplets, whereas for $N$ odd we find a nonlagrangian theory. The case $N=5$, whose manifest symmetry in this setup is $SO(10)\times U(1)$, corresponds to $D_2(SO(8))$, which was identified in \cite{Cecotti:2013lda} with the $E_6$ Minahan-Nemeschansky theory \cite{Minahan:1996fg}. 

Let's now consider exceptional theories with $k=1$. As already explained, the case $b=J$ is trivial so we are left with five nontrivial models we will now analyze\footnote{Three of these models are also discussed in \cite{Song:2017oew}.}: 
\begin{itemize}
\item The theory $D_1^9(E_6)$ has a one dimensional Coulomb branch generated by an operator of dimension 3 (corresponding to the deformation $u_3t$) and no mass parameters except the Casimirs of $E_6$. The only theory with $E_6$ global symmetry and a Coulomb branch of this form is the $E_6$ Minahan-Nemeschansky theory. 
\item $D_1^8(E_6)$ has one Coulomb branch coordinate (the deformation is $u_4t$) of dimension four and one mass parameter. We conclude that the theory has (at least) $U(1)\times E_6$ global symmetry and a CB operator of dimension 4, so we identify it with $E_7$ Minahan-Nemeschansky theory \cite{Minahan:1996cj}. 
\item $D_1^{14}(E_7)$ has again only one CB operator (the deformation is $u_4t$) of dimension four and $E_7$ global symmetry, so we are led to identify it with $E_7$ Minahan-Nemeschansky theory. 
\item The Coulomb branch operator of $D_1^{24}(E_8)$ (which still has rank one) has dimension 6 (the deformation is $u_6t$) therefore we identify it with $E_8$ Minahan-Nemeschansky theory \cite{Minahan:1996cj}. 
\item Finally, $D_1^{20}(E_8)$ has two Coulomb branch operators of dimension 4 and 10 respectively (the corresponding deformations are $u_4x_3t$ and $u_{10}t$). Since we do not find any mass parameter except the $E_8$ Casimirs we conclude that the global symmetry is exactly $E_8$. Using the technology of the previous section we find that the flavor central charge is 20, the a and c central charges are $a=101/12$ and $c=31/3$. This model arises in the $E_7$ tinkertoys classification \cite{Chacaltana:2017boe} as one factor inside trinions corresponding to product SCFTs\footnote{I would like to thank Jacques Distler for pointing this out.}. We find that upon turning on a principal nilpotent vev for the $E_8$ global symmetry this model flows in the IR to $A_4$ AD theory. This is a new prediction of our construction.
\end{itemize} 
Notice that the value of the flavor central charge for $D_k^b(J)$ theories (\ref{betacon}) and also the $a,c$ central charges are perfectly consistent with all the identifications proposed above. 

We would now like to comment about the $J=SU(2)$ case. It is known that $D_k(SU(2))$ is equivalent $(A_1,D_k)$ theory; what about the $b=1$ case? As we have seen before, the $D_k^1(SU(2))$ model is described by the SW curve $x^2+xt^k=0$. Modulo a shift of $x$ which does not affect the differential (always up to exact terms) the curve can be brought to the form $x^2+t^{2k}=0$, which is just the SW curve for $D_{2k}(SU(2))=(A_1,D_{2k})$. Again the values of $a$, $c$ and the flavor central charge are consistent with this claim. We thus simply get a subclass of $D_k(SU(2))$ theories.  

Let us now move to the lagrangian subclass. We start by noticing that there are no lagrangian theories for $J=E_N$: this was already established for $D_k(E_N)$ theories in \cite{Cecotti:2013lda} and the same argument used there rules out lagrangian theories in the other cases as well. The contribution $\beta$ of $D_k^b(J)$ theories to the $J$ beta function is always less than $h(J)$ (see (\ref{betacon})), so none of the $J=E_8$ theories can be lagrangian since all matter fields in a nontrivial representation of $E_8$ contribute at least $2h(J)$. In the $E_7$ and $E_6$ cases the only allowed matter fields compatible with the above constraint are half-hypermultiplets in the {\bf 56} of $E_7$ and full hypermultiplets in the {\bf 27} of $E_6$. In both cases the contribution to the beta function is 6. From table (\ref{sing}) it is easy to see that there are no values of $k$ and $b\neq h(J)$ such that (\ref{betacon}) is a multiple of 6. 

Let us now consider the more interesting case of classical Lie groups ($J=A_{N-1}$ and $J=D_N$). One obvious lagrangian subclass is $D_1^{N-1}(SU(N))$ which just describes $N$ free hypermultiplets\footnote{Indeed this class of theories does not display any interesting dynamics, however it can be considered an example of supersymmetry enhancement in its own right: when we couple the chiral multiplet in the adjoint of $SU(N)$ and give it a vev, $N-1$ hypermultiplets become massive and only one of them survives at low energy. This is the IR fixed point. Our geometric setup is perfectly consistent with this conclusion: the CY we associate with the IR theory is $$x_1^2+x_2^2+x_3^N+x_3z=0.$$ With the redefinition $z'=z+x_3^{N-1}$ this becomes 
$$x_1^2+x_2^2+x_3z'=0$$ which is known to describe a single hypermultiplet.}. A more interesting lagrangian subclass is provided by $D_k(J)$ theories. This was already discussed in \cite{Cecotti:2013lda} so we will simply state the result: 
\begin{itemize}
 \item For $J=SU(N)$ the model is lagrangian iff $N$ is a multiple of $k$ and in this case the theory is the following linear quiver of special unitary groups:
\be\label{quiver3} SU(n)-SU(2n)-SU(3n)-\dots -SU(N-n)-\boxed{N}\ee
The quiver terminates with $N$ fundamentals of $SU(N-n)$ ($n=N/k$). 
\item For $J=SO(2N)$ the theory is lagrangian only if $N=nk+1$ for an arbitrary positive integer $n$. These models correspond to linear quivers of alternating SO/USp gauge groups with half-hypermultiplets in the bifundamental representation. There are two cases depending on the parity of $k$. For $k$ odd the theory is 
\be\label{quiver1} SO(2n+2)-USp(4n)-\dots-USp(2N-2n-2)-\boxed{N}\ee 
and for $k$ even we have 
\be \boxed{1}-USp(2n)-SO(4n+2)-\dots-USp(2N-2n-2)-\boxed{N}\ee
\end{itemize}

Let's now consider the remaining cases: $D_k^{N-1}(SU(N))$ and $D_k^{N}(SO(2N))$. For $D_k^{N-1}(SU(N))$ theories the contribution to the $SU(N)$ beta function is $\beta=N-\frac{N-1}{k}$. Of course the theory can be lagrangian only if $\beta$ is an integer, which demands $N=kn+1$. In this case the Coulomb branch spectrum and Seiberg-Witten curves discussed before agree with those of the following linear quiver (see \cite{Witten:1997sc}): 
\be\label{quiver4}\boxed{1}-SU(n+1)-SU(2n+1)-\dots-SU(N-n)-\boxed{N}\ee 
so we are led to identify $D_k^{N-1}(SU(N=kn+1))$ with the above linear quiver. 

In the case of $D_k^{N}(SO(2N))$ theories $\beta=2N-2-\frac{N}{k}$ and the model can be lagrangian only if $\beta$ is an even integer, implying $N=2nk$ for some positive integer $n$. We claim that this class of models coincides with the following family of linear quivers (the number of gauge groups is $2k-1$):
\be\label{quiver2}USp(2n-2)-SO(4n)-\dots-SO(2N-4n)-USp(2N-2n-2)-\boxed{N}\ee 
Indeed the SW curve found in \cite{Landsteiner:1997vd} for this quiver is identical to (\ref{quiverso2}).

\section{Comparison with the literature, new examples of susy enhancement, SW curves and chiral algebras}

\subsection{Recovering all lagrangians for nonlagrangian theories}

Having identified the lagrangian subclass, we can now notice that the lagrangian theories for which the Maruyoshi-Song procedure is known to ``work'' (meaning that supersymmetry enhances in the infrared) are precisely those listed in section \ref{lagrth} (with the exception of the $D_1^{N-1}(SU(N))$ free theories)! Let us then check that the duality we are proposing is consistent with the results available in the literature: 
\begin{itemize}
 \item According to our duality the quiver theory $D_k(SU(nk))$ flows under the MS procedure to the $SU_{nk}^{nk}(k)$ theory, which is the same as $(A_{nk-1},A_{k-1})$ theory. This is precisely the answer found in \cite{Benvenuti:2017bpg, Agarwal:2017roi}. In the special case $k=2$ we recover the observation of \cite{Maruyoshi:2016aim} that conformal $SU(n)$ SQCD flows in the IR to the $A_{2n-1}$ AD theory when we turn on a principal nilpotent vev. 
 \item The prediction of the duality is that the two ortho-simplectic quivers corresponding to the $D_k(SO(2nk+2))$ theory with k even or odd flows in the IR to the $SO(2nk+2)^{2nk+2}(k)$ theory, which coincides with the $(A_{k-1},D_{nk+1})$ theory. Indeed this is the answer found in \cite{Agarwal:2017roi}. As a special case we recover for $k=2$ the statement that under a next-to-maximal nilpotent vev $USp(2n)$ conformal SQCD flows to the $D_{2n+1}$ theory \cite{Agarwal:2016pjo}. 
 \item The next prediction is that the unitary quiver $D_k^{kn}(SU(kn+1))$ flows in the IR to the model $SU(kn+1)^{kn}(k)$ which is the same as the $(I_{k,kn},S)$ theory. This was checked at the level of central charges in \cite{Agarwal:2017roi} and also at the level of the 3d mirror in \cite{Benvenuti:2017bpg}. For $k=2$ this reduces to the statement that under a next-to-maximal nilpotent vev $SU(n)$ conformal SQCD flows to the $D_{2n}$ theory \cite{Agarwal:2016pjo}. 
 \item Finally, we have the quiver $D_k^{2nk}(SO(4nk))$ which was also discussed in \cite{Agarwal:2017roi}. Our duality predicts that the IR fixed point is the $SO(4nk)^{2nk}(k)$ theory. One can readily check using the technology reviewed in this paper that the Coulomb Branch and the $a$, $c$ central charges of the IR SCFT match perfectly those of the $SO(4nk)^{2nk}(k)$ theory. For $k=1$ we find the lagrangian UV completion of $(A_1,A_{2n-2})$ theories studied in \cite{Maruyoshi:2016aim}. Further specializing to the case $n=2$ we recover the result of \cite{Maruyoshi:2016tqk}. 
 \end{itemize} 
 
 \subsection{Non principal nilpotent vevs and new examples of susy enhancing RG flows}
 
 In \cite{Agarwal:2016pjo} the authors found several examples of theories which exhibit supersymmetry enhancement in the IR upon turning on a non principal nilpotent vev. As was stated in the introduction, our claim is that in our geometric framework focusing on principal nilpotent vevs is not restrictive and we recover anyway all the susy enhancing RG flows. This effectively makes the choice of nilpotent vev, which is part of the defining data of the Maruyoshi-Song construction, redundant as long as one is interested in susy enhancing RG flows only. Let us then check that the results of \cite{Agarwal:2016pjo} are reproduced by our procedure. 
 
 The discussion of the previous subsection already includes some examples: it is known that in the case of $SU(N)$ SQCD with $2N$ flavors there are two choices of nilpotent vevs (principal and subregular) which lead to supersymmetry enhancement in the IR. Accordingly, we have two different realizations of conformal $SU(N)$ SQCD in our class: $D_2(SU(2N))$ and $D_2^{2N-2}(SU(2N-1))$. Considering the principal nilpotent vev for $SU(2N-1)$, which is the manifest global symmetry $J$ in the second realization, is just equivalent to considering the subregular nilpotent vev for the full $SU(2N)$ global symmetry, so in this sense we do not miss the subregular case in our setup. Analogously, the two possible choices of nilpotent vev for $USp(2N)$ conformal SQCD, whose global symmetry is $SO(4N+4)$, correspond to two different realizations of this theory in the $D_k^b(J)$ class: $D_1^{2N+2}(SO(4N+4))$ and $D_2(SO(4N+2))$. 
 
 Another simple example is $D_4$ AD theory, which has $SU(3)$ global symmetry and flows to $\mathcal{N}=2$ SCFTs under both choices of nilpotent vev (principal and minimal). Again, this is reproduced by focusing on the principal nilpotent vev for the manifest symmetry in the geometric description: $D_4$ AD theory is equivalent to either $D_2(SU(3))$ or $D_2^1(SU(2))$. More in general, it was pointed out in Section \ref{lagrth} that $D_2(SU(N+1))$ and $D_2^{N-1}(SU(N))$ are equivalent, although the full $SU(N+1)$ global symmetry of the theory is not manifest in the geometric setup. Our duality then predicts that $D_2(SU(N+1))\simeq (I_{N+1,1-N},F)$ exhibits susy enhancement under both principal and subregular nilpotent vevs and the IR fixed points are respectively $A_N$ and $D_{N+1}$ AD theories. This is in perfect agreement with the findings of \cite{Agarwal:2016pjo}. All other choices of nilpotent vev do not lead to supersymmetry enhancement. 
 
 Finally, the case of $E_6$ Minahan-Nemeschansky theory is particularly interesting: in \cite{Agarwal:2016pjo} the authors examined all possible nilpotent vevs and concluded that the scaling dimension of operators at the IR fixed point are always irrational except in three cases which exhibit supersymmetry enhancements. In these distinguished cases the $SU(2)$ group defining the nilpotent orbit is embedded in a $SO(8)$ or $SO(10)$ subgroup of $E_6$ (the third case is just the principal nilpotent orbit which has no commutant). 
 As we have seen in the previous section, $E_6$ Minahan-Nemeschansky theory appears three times in the $D_k^b(J)$ class: it is equivalent to $D_2(SO(8))$, $D_1^{5}(SO(10))$ and $D_1^9(E_6)$. Depending on the specific realization only a subgroup of the $E_6$ symmetry is explicit (precisely the three subgroups mentioned before). According to our duality, by turning on a principal nilpotent vev for the $J$ subgroup we recover $\CN=2$ supersymmetry in the infrared. The predicted IR fixed points are respectively $SO(8)^6(2)=(A_1,D_4)$, $SO(10)^5(1)=(A_1,D_3)$ and $E_6^9(1)=(A_1,A_2)$ (these equivalences can be easily derived from the geometric engineering setup discussed before). These models coincide precisely with those found in \cite{Agarwal:2016pjo}. It is very satisfactory to see that our framework automatically selects all the nilpotent orbits of $E_6$ which lead to supersymmetry enhancement. This fact strongly suggests this is the right framework to understand the susy enhancing RG flow. We would also like to notice that in Section \ref{lagrth} we found realizations of $E_7$ and $E_8$ Minahan-Nemeschansky theories with full manifest global symmetry: $E_1^{14}(E_7)$ and $E_1^{24}(E_8)$. The corresponding IR fixed points are in both cases equivalent to $A_2$ AD theory in agreement with the findings of \cite{Maruyoshi:2016aim}. We also find a second realization of $E_7$ Minahan-Nemeschansky ($D_1^{8}(E_6)$) in which the manifest global symmetry is $E_6$. This tells us the theory exhibits enhanced supersymmetry under a non principal nilpotent vev and the IR fixed point is $E_6^{8}(1)$, which can be shown to be equivalent to $A_3$ AD theory using the geometric engineering technology reviewed above. This RG flow has not been noticed before. 
  
 As we have just seen, all the RG flows with supersymmetry enhancement discussed in \cite{Maruyoshi:2016tqk,Maruyoshi:2016aim,Agarwal:2016pjo,Benvenuti:2017bpg, Agarwal:2017roi} fit in our framework and our construction predicts a new (non lagrangian) example involving $E_7$ Minahan-Nemeschansky theory. This is just one out of infinitely many new such flows: all the flows involving $D_k^b(E_N)$ theories with $k>1$ and non lagrangian $D_k^b(SO(2N))$, $D_k^{N-1}(SU(N))$ models have not been discussed before and represent a prediction of our construction. Below we will describe in detail other examples.  
  
  In \cite{Agarwal:2016pjo} it was found that when the nilpotent vev leaves a subgroup of the global symmetry unbroken, usually there is no SUSY enhancement in the IR. Sometimes this can be ``remedied'' by gauging the surviving global symmetry. For instance, if we consider $E_8$ MN theory and turn on a principal nilpotent vev for a $E_6$ subgroup with commutant $SU(3)$ (inside $E_8$) we get irrational R-charges at the IR fixed point. However, if we gauge this $SU(3)$\footnote{The embedding index of the $E_6$ subgroup is one, hence the gauging is conformal.} the conclusion changes and the theory flows in the IR to $E_6$ AD theory. This fits perfectly in our setup: $D_2(E_6)$ is precisely an $SU(3)$ gauging of $E_8$ MN. 
 
 One further example is provided by $D_2^{24}(E_8)$. This model has CB operators of dimension $(2,6,6,8,12,18)$, $\beta_{E_8}=18$ and central charges $a=\frac{111}{4}$, $c=31$. Since the spectrum includes a dimension 2 CB operator, we know there is an exactly marginal coupling \cite{Buican:2014hfa}. We identify this with the gauge coupling of a $G_2$ vectormultiplet. Our proposal implies that the ``matter sector'' of $D_2^{24}(E_8)$ is a rank four theory with CB operators of dimension $(6,8,12,18)$. The flavor symmetry is $E_8\times G_2$ and the $G_2$ flavor central charge has to be $4h^\vee(G_2)=16$ for the gauging to be conformal. We can also compute $a$ and $c$ central charges just by subtracting the contribution of $dim(G_2)=14$ vectormultiplets. This leads to $a=\frac{149}{6}$ and $c=\frac{86}{3}$. Indeed a theory with exactly these properties is already known \cite{Chacaltana:2017boe}\footnote{The value of the central charges can be found following the procedure of \cite{Ohmori:2015pua} from the anomaly polynomial of the six dimensional theory \cite{Ohmori:2014kda}.}: it is the ($T^2$ compactification of) $(E_8,G_2)$ conformal matter \cite{DelZotto:2014hpa}. This also arises as a trinion in the $E_7$ tinkertoys classification. The corresponding IR fixed point under the susy enhancing RG flow is the $Q_{12}$ model discussed in \cite{Cecotti:2011gu}. If instead we consider the susy enhancing RG flow for the $E_8$ symmetry of the (4d) $(E_8,G_2)$ conformal matter alone, we find that the infrared scaling dimensions are irrational.

\subsection{Seiberg-Witten curves and the susy enhancing RG flow}

There are at least two interesting outcomes of our construction. The first is a direct way to relate the SW data (curve and differential) of the UV and IR fixed points of the susy enhancing RG flow (at least in the $J=A,D$ cases, which anyway include all lagrangian UV completions of AD theories). Our observation is a direct consequence of the fact that $D_k^b(J)$ theories (UV fixed point) and $J^b(k)$ theories (IR fixed point) are described by the same hypersurface $W(x_{1,2,3},t)=0$ (in $\mathbb{C}^3\times\mathbb{C}^*$ and $\mathbb{C}^4$ respectively) and the holomorphic three-forms are respectively (see (\ref{normal})) $\prod_idx_idt/(tdW)$ and $\prod_idx_idt/dW$.

The procedure is rather simple to state: for every $D_k^b(J)$ theory consider the SW curve and differential, which has the canonical form $xdt/t$. The curve describing the IR fixed point is the same, and the correct SW differential is obtained just by dropping the denominator (hence it has the simple form $xdt$). The only subtlety is that the set of allowed deformations (i.e. CB coordinates and mass parameters) of the singular curve changes, so the singular curves always agree but the fully deformed curves may differ. This issue can be handled by going back to the CY geometry (\ref{sing}) and identifying all allowed deformations, which are encoded in the quotient algebra $\mathcal{A}_W$ as explained in the previous sections. A small caveat is that sometimes the presentation of the SW curve for the IR theory provided by our algorithm is not the conventional one.

Let us see some examples to illustrate the procedure. The SW curve of $D_k(SU(N))$ theories is $x^N+t^k=0$. The same curve but with SW differential $xdt$ describes $(A_{N-1},A_{k-1})$ theories, which are precisely the resulting IR fixed points under the susy enhancing RG flow. Similarly, the family of curves $x^N+xt^k=0$ describes $D_k^{N-1}(SU(N))$ theories if the SW differential is $xdt/t$. On the other hand, when $\lambda_{SW}=xdt$ these correspond to the models one gets by compactifying the $A_{N-1}$ $\CN=(2,0)$ theory on a sphere with one irregular puncture of type II, in the notation of \cite{Xie:2012hs}. As was pointed out in the same paper, these are equivalent to $(I_{k,N-1},S)$ theories. This statement can be readily checked by writing the curve in the standard class $\mathcal{S}$ form, where $x$ is now interpreted as the coordinate parametrizing the UV curve and $t$ the coordinate on the fiber of the cotangent bundle. 

A slightly more involved example is given by $D_1^N(SO(2N))$ theories, which flow under the MS flow to $A_{N-2}$ AD theories. For $N=2n$ their SW curve takes the form \be\label{swann}x^{2N}+\sum_{i=1}^{N-1}m_ix^{2i}+x^2P_{n-1}(x^2)t+(t+m)^2=0\quad\lambda_{SW}=x\frac{dt}{t},\ee 
where $P_{n-1}$ is a generic polynomial of degree $n-1$. We derived this formula from (\ref{quiverso2}), keeping all complex structure deformations turned on. According to our claim, by replacing the SW differential with $xdt$ we get the SW data associated with the IR fixed point (in this case $A_{N-2}$ AD theory). At first sight the curve (\ref{swann}) does not look like the more familiar expression 
\be\label{swant}y^2=x^{N-1}+\sum_{i=0}^{N-3}u_ix^{i}\quad\lambda_{SW}=ydx,\ee 
but a change of variables relates the two: start from (\ref{swann}) with SW differential $xdt=-tdx$ and redefine $t'=t+m$. Up to exact terms the SW differential retains the canonical form $t'dx$\footnote{Notice that in the UV theory this shift would change the location of the pole for the SW differential so is not as harmless as in this case.} and the parameter $m$ drops out from the curve since it can be reabsorbed with a redefinition of the $m_i$ parameters. With the further redefinition $\tilde{t}=-t'-(x^2P_{n-1}(x^2))/2$ (\ref{swann}) becomes 
$$\tilde{t}^2+x^2Q_{N-1}(x^2)=0 \quad\lambda_{SW}=\tilde{t}dx,$$ 
where $Q_{N-1}$ is a generic polynomial of degree $N-1$. Introduce now $y=\tilde{t}/x$ so that the SW differential becomes $yd(x^2)/2$. If we now divide the curve by $x^2$ and set $x'=x^2$, we find that (\ref{swann}) reduces precisely to (\ref{swant}) modulo shifting $x'$. The case $N=2n+1$ is analogous.

\subsection{Comments about chiral algebras}

The second implication is a connection at the level of chiral algebras \cite{Beem:2013sza}, at least for $D_k^b(J)$ theories whose global symmetry is exactly $J$, without any enhancement (these include some of the lagrangian theories discussed before, specifically the linear quivers (\ref{quiver1}) and (\ref{quiver2})) It was conjectured in \cite{Xie:2016evu} that the chiral algebra for this class of models is given by the affine Kac-Moody (AKM) algebra of type $J$ at level $-\beta=\frac{b}{k}-h(J)$ (denoted also as $\widehat{J}_{-\beta}$). Since the susy enhancing procedure breaks extended supersymmetry, it is not obvious how to describe the RG flow at the level of chiral algebras. In this sense our duality comes to the rescue since the problem can be circumvented by looking at the dual frame, which is manifestly $\CN=2$ preserving. First of all we go from $D_k^b(J)$ to the UV fixed point of the dual RG flow, the $D_{k+b}^b(J)$ theory, whose chiral algebra is again (assuming the conjecture of \cite{Xie:2016evu}) AKM of type $J$ but now with level $\frac{b}{k+b}-h(J)$. 

Notice that the rank of the global symmetry $r(G_F)$ of $D_k^b(J)$ theories for given $J$ depends only on $b$ and on $k(modb)$ (see Table (\ref{massnum}) in the Appendix). In particular, $r(G_F)$ is the same for $D_k^b(J)$ and $D_{k+b}^b(J)$ theories. This property is essential for our argument because the chiral algebra is AKM only when the rank of the global symmetry is $r(J)$ and the shift in $k$ dictated by our duality does not take us out of this subclass. The IR fixed point is then obtained by closing the full puncture and this procedure corresponds to quantum Drinfel'd-Sokolov (qDS) reduction at the level of chiral algebras (see \cite{Beem:2013sza}). Combining the two operations, we conclude that the net effect of the susy enhancing RG flow is the following: we should shift the level of the AKM algebra as stated above and then perform qDS reduction.  

This picture needs to be refined for $D_k^b(J)$ theories with enhanced global symmetry (the simplest examples are $D_{2n}(SU(2))=(A_1,D_{2n})$ theories and the unitary quivers (\ref{quiver3}), (\ref{quiver4})), since the chiral algebra is no longer AKM: in this case there are extra Higgs chiral ring generators which always correspond to strong generators of the chiral algebra \cite{Beem:2013sza} (see \cite{Beem:2017ooy} for the relation between chiral algebras and the Higgs Branch of the underlying $\CN=2$ theory), so we conclude that for this class of theories the affine algebra of type $J$ at level $-h(J)+\frac{b}{k}$ can only be a subalgebra (this is indeed the case for conformal $SU(N)$ SQCD \cite{Beem:2013sza} and for $D_{2n}$ AD theories \cite{Creutzig:2017qyf}). Because of this fact, it is not obvious how to relate the chiral algebras of $D_k^b(J)$ and $D_{k+b}^b(J)$ theories in general. Understanding the Higgs branch of these models would be for sure helpful to shed light on this problem, but we leave this for future research. 

In some special cases the problem can be circumvented as follows: many $D_k^b(J)$ theories admit several class $\mathcal{S}$ realizations and sometimes this can be exploited to identify the corresponding chiral algebra as in \cite{Song:2017oew}. Let us discuss the case of $D_{2n}$ Argyres-Douglas theories to illustrate this point: the observation of \cite{Song:2017oew} is that these models can be obtained starting from $D_{n}(SU(n+1))$ and turning on a nilpotent vev for the $SU(n+1)$ moment map with a Jordan block of size $n-1$ (labelled by the partition $(n-1,1,1)$). With a further nilpotent vev for the surviving $SU(2)$ symmetry we flow to $A_{2n-3}$ AD theory. More directly, we get $A_{2n-3}$ AD by turning on the nilpotent vev labelled by the partition $(n-1,2)$. The key observation now is that the global symmetry of $D_{n}(SU(n+1))$ is exactly $SU(n+1)$ with no enhancement, so the corresponding chiral algebra is 
$SU(n+1)$ AKM at level $\frac{1-n^2}{n}$. We can then obtain the chiral algebras associated with $D_{2n}$ and $A_{2n-3}$ via qDS reduction, in agreement with \cite{Creutzig:2017qyf}. Notice that for $n=2$ we recover the known result that the chiral algebra of $D_4$ AD is $\widehat{SU(3)}_{-\frac{3}{2}}$ (see \cite{Buican:2015ina,Cordova:2015nma,Rastelli}). Exploiting this observation we see immediately that a possible extension of our costruction to $D_{2n}$ theories (so we have a precise map for all $D_k^b(SU(2))$ theories) is the following: consider $SU(n+1)$ AKM at level $\frac{1-n^2}{n}$ (whose $(n-1,1,1)$\footnote{We use the partition labelling the nilpotent orbit to specify the qDS reduction we need.} qDS reduction gives $D_{2n}$ AD), increase $n$ by one unit and then perform qDS reduction associated with the nilpotent orbit labelled by the partition $(n,2)$. Presumably a construction along these lines works in several other cases. 

\section{Concluding remarks}

In this paper we have given evidence that geometric engineering in Type IIB is the right framework to study susy enhancing RG flows. This realization allows to identify immediately the IR fixed point, without need to perform a-maximization, and provides a simple relation between the SW geometry of UV and IR fixed points. All known examples of supersymmetry enhancement (in particular lagrangian theories) fit in our framework and in this paper we find several new nonlagrangian examples from our duality. This approach represents a more systematic treatment of susy enhancing RG flows and provides the (so far) missing pattern underlying the various examples discussed in the literature. The analysis is also simplified due to the fact that this setup directly singles out the ``distinguished'' choices of nilpotent vevs which lead to supersymmetry enhancement in the infrared, effectively making this extra input unnecessary. This is due to the existence of multiple geometric realizations of the same superconformal theory.

A key point is the extremely simple relation between the geometric descriptions of the UV and IR fixed points: the two theories are simply defined compactifying Type IIB on hypersurfaces described by the same equation in $\mathbb{C}^3\times\mathbb{C}^*$ and $\mathbb{C}^4$ respectively. This simplicity makes some general properties of the RG flow more manifest. This fact remains an empirical observation in the present work and it would be interesting to understand better its origin. In particular, it would be important to understand how to describe in Type IIB the $\mathcal{N}=1$ deformation we have discussed and achieve a geometric description of the whole RG flow. 

The geometric relation between UV and IR theories can also be formulated in the class $\mathcal{S}$ language, although it becomes more involved in that setup: the UV theory is described by a sphere with one irregular puncture and a full regular puncture. The IR theory is then obtained by closing the regular puncture and increasing the order of the pole of the Hitchin field at the irregular puncture by one unit. This link with the class $\mathcal{S}$ construction suggests some natural generalizations of our work: one could try for example to incorporate irregular punctures of Type III (in the notation of \cite{Xie:2012hs}), which provide already in the $J=A_N$ case a large class of models which do not fit in our framework (see for instance \cite{Buican:2014hfa,Buican:2017fiq}), or twisted irregular punctures. 

There are several other directions worth exploring: first of all it would be important to prove our duality (we mean deriving it from known infrared dualities). This would explain why Argyres-Douglas theories appear when we apply the susy enhancing procedure to $\mathcal{N}=2$ gauge theories. Another interesting generalization is to find lagrangian UV completions for $D_k^b(J)$ theories. This would enlarge significantly the lanscape of UV lagrangians for strongly-coupled $\mathcal{N}=2$ theories. Their existence is not unreasonable since we already have some examples: all $D_n$ AD theories are in the $D_k^b(J)$ class and a UV lagrangian completion for those is known. Finally, an interesting observation is that the susy enhancing RG flows always preserve the dimension of the Coulomb branch. As we have remarked in Section \ref{dualsec}, this feature has interesting implications and deserves further attention. In particular, it would be important to achieve a field-theoretic derivation of it. We hope to come back to these points in the near future.

\section*{Acknowledgments} 
I would like to thank Matthew Buican and Noppadol Mekareeya for carefully reading the manuscript and for their suggestions. I also acknowledge Sergio Benvenuti, Sergio Cecotti, Jacques Distler and Philip Argyres for useful discussions. This project is supported by ICTP and by the INFN Research Project ST\&FI.

\appendix

\section{RG flows, a-maximization and higgsing} 

In this section we explain how to compute central charges and scaling dimensions of chiral operators at the IR fixed point of the RG flows we consider in this paper. We consider first the susy enhancing RG flow following the analysis given in \cite{Gadde:2013fma} (see also \cite{Agarwal:2014rua}) and then the closure of the full puncture, which is discussed in detail in \cite{Tachikawa:2015bga}.

\subsection{Susy enhancing RG flow}

The R-symmetry at the infrared fixed point can be determined as follows: first of all we exploit the fact that every $\CN=2$ superconformal theory has two canonical $U(1)$ global symmetries (the $U(1)_R$ group $R_{\CN=2}$ and the cartan of the $SU(2)_R$ symmetry $I_3$). If the theory has global symmetry $J$, when we add a chiral multiplet $M$ transforming in the adjoint of $J$ and give it a principal nilpotent vev, we should also consider the Cartan $\rho(\sigma_3)$ of the $SU(2)$ embedding labelling the nilpotent orbit (our convention is $\langle M\rangle=\rho(\sigma^+)$). Out of these three $U(1)$'s one combination is broken by the vev and we assume the $U(1)$ R-symmetry of the IR fixed point is a combination of the surviving two, which we can parametrize as follows: 
\be\label{rtrial}R_{\epsilon}=\frac{1+\epsilon}{2}R_{\CN=2}+(1-\epsilon)I_3-(1+\epsilon)\rho(\sigma_3).\ee 
The value of $\epsilon$ can be determined via a-maximization applying the following procedure. We start by computing the trial 
central charges 
\be\label{amax} a(\epsilon)=\frac{3}{32}(3\Tr R_{\epsilon}^3-\Tr R_{\epsilon});\quad c(\epsilon)=a(\epsilon)-\frac{\Tr R_{\epsilon}}{16}.\ee
Plugging in (\ref{rtrial}) we find the expression
\be\label{atr1} a(\epsilon)=\frac{3}{32}\left[\frac{3}{8}(1+\epsilon)^3\Tr R_{\CN=2}^3+\frac{9+9\epsilon}{2}[(1-\epsilon)^2\Tr R_{\CN=2}I_3^2-(1+\epsilon)^2I_{\rho}\beta]-\frac{1+\epsilon}{2}\Tr R_{\CN=2}\right],\ee 
\be c(\epsilon)=a(\epsilon)-\frac{1+\epsilon}{32}\Tr R_{\CN=2}.\ee
In (\ref{atr1}) $\beta$ denotes the $J$ flavor central charge divided by two (see (\ref{betacon})) and $I_{\rho}$ is the embedding index of the $U(1)$ group generated by $\rho(\sigma_3)$ inside $J$. For the principal nilpotent orbit of a simply-laced group (which is the only case we need in the present paper) the embedding index is \cite{Embindex} 
\be\label{embind}I_{\rho}=\frac{h(J)Dim(J)}{6},\ee 
where $Dim(J)$ is the dimension of the group $J$. 
Using now the well-known formulas for $\CN=2$ SCFT's 
\be \Tr R_{\CN=2}^3=\Tr R_{\CN=2}=48(a-c);\quad \Tr R_{\CN=2}I_3^2=4a-2c.\ee 
(all other 't Hooft anomalies are trivial) we can rewrite the trial central charges in terms of $\beta$, $I_{\rho}$ and the a,c central charges of the $D_k^b(J)$ theory.

When we turn on the vev, some components (precisely $Dim(J)-r(J)$ of them) of the chiral multiplet $M$ decouple and are identified with the Goldstone multiplets of the spontaneous symmetry breaking. Consequently, in order to extract the central charges of the interacting sector of the IR fixed point, we need to add the contribution from the gauge singlets $M_i$ which do not decouple from the theory. These have trial R-charge $C_i(J)(1+\epsilon)$, where $C_i(J)$ denote the degree of the Casimir invariants of $J$. Their contribution to the trial $a$ and $c$ central charges is: 
\begin{itemize}
\item For $J=SU(N)$
\be a'=\frac{3}{128}(N-1)(N+\epsilon(N+2))(6\epsilon^2-2+3N^2(1+\epsilon)^2+3N(\epsilon^2-1)),\ee 
\be c'=a'-\frac{N^2-N}{32}-\epsilon\frac{N^2+N-2}{32}.\ee 
\item For $J=SO(2N)$ 
\be a'=\frac{3}{32}N(N-1+N\epsilon) (8+6N^2(1+\epsilon)^2 + 6\epsilon(2+\epsilon)-3N(1+\epsilon)(5+3\epsilon)),\ee
\be c'=a'-\frac{1}{16}(N^2-N+N^2\epsilon).\ee
\item For $J=E_6$ 
\be a'=\frac{9}{16}(6+7\epsilon)(197+3\epsilon(144+79\epsilon));\quad c'=a'-\frac{1}{8}(18+21\epsilon).\ee 
\item For $J=E_7$ 
\be a'=\frac{21}{16}(9+10\epsilon)(229+6\epsilon(81+43\epsilon));\quad c'=a'-\frac{1}{16}(63+70\epsilon).\ee 
\item For $J=E_8$ 
\be a'=\frac{3}{2}(15+16\epsilon)(652+3\epsilon(450+233\epsilon));\quad c'=a'-\frac{1}{2}(15+16\epsilon).\ee
\end{itemize}

By examining a large number of examples we find that, once all the operators violating the unitarity bound are decoupled (or flipped according to our prescription), the trial a central charge is maximized at 
\be\label{epsmax} \epsilon=-\frac{k+3b}{3k+3b}.\ee 
Notice that the singlet of R-charge $h(J)(1+\epsilon)$ becomes in the IR the CB operator with largest scaling dimension. Using (\ref{epsmax}) we find that its dimension is $\frac{h(J)k}{k+b}$, which agrees precisely with the result found in section 2 for the model $J^b(k)$. 

\subsection{Closure of the full puncture}

If instead we are interested in higgsing the theory by turning on a principal nilpotent vev for the $J$ moment map $\mu$ (again our convention for the vev is $\langle\mu\rangle=\rho(\sigma^+)$), the trial R-charge can be written as follows (the notation is identical to the susy enhancing case): 
\be\label{rtrial2} R_{\epsilon}=\frac{1+\epsilon}{2}R_{\CN=2}+(1-\epsilon)I_3-(1-\epsilon)\rho(\sigma_3).\ee 
The trial central charges then read 
\be\label{higgs2} a(\epsilon)=\frac{3}{32}\left[\frac{3}{8}(1+\epsilon)^3\Tr R_{\CN=2}^3+\frac{9}{2}(1+\epsilon)(1-\epsilon)^2[\Tr R_{\CN=2}I_3^2-I_{\rho}\beta]-\frac{1+\epsilon}{2}\Tr R_{\CN=2}\right],\ee 
\be c(\epsilon)=a(\epsilon)-\frac{1+\epsilon}{32}\Tr R_{\CN=2}.\ee
In order to isolate the information about the interacting sector, we should now subtract by hand the contribution from the Goldstone multiplets (see \cite{Tachikawa:2015bga}). In order to explain how this is done, we should remind the reader that the $J$ moment map (which indeed transforms in the adjoint representation) decomposes into the direct sum of $r(J)$ irreducible representations of the $SU(2)$ subgroup labelling the principal nilpotent orbit and their spin is equal to the set of exponents $E_i(J)$ (degree of the Casimir invariants minus one) of the Lie algebra. All the components of the moment map except the lowest weight state of each $SU(2)$ irrep are Goldstone multiplets (see section 2.4 of \cite{Tachikawa:2015bga}) and there are $Dim(J)-r(J)$ of them, exactly as in the susy enhancing case. Their charge under (\ref{rtrial2}) is 
\be(1+n-E_i(J))(1-\epsilon)\quad 0\leq n\leq 2E_i(J)-1.\ee

Plugging this inside (\ref{amax}) we find the contribution from Goldstone multiplets to the trial central charges: 
\begin{itemize} 
 \item For $J=SU(N)$ 
 \be\tilde{a}=-\frac{3}{128}N(N-1)(1+\epsilon)(-2+3(N-1)N(\epsilon-1)^2+6\epsilon),\ee 
 \be\tilde{c}=\tilde{a}+\frac{N(N-1)}{32}(1+\epsilon).\ee
\item For $J=SO(2N)$ 
\be\tilde{a}=-\frac{3}{32}(N-1)N(1+\epsilon)(8-15N(\epsilon-1)^2+6N^2(\epsilon-1)^2+3\epsilon(3\epsilon-5)),\ee
\be\tilde{c}=\tilde{a}+\frac{N(N-1)}{16}(1+\epsilon).\ee 
\item For $J=E_6$ 
\be\tilde{a}=-\frac{27}{8}(1+\epsilon)(197+3\epsilon(66\epsilon-131));\quad \tilde{c}=\tilde{a}+\frac{9}{4}(1+\epsilon).\ee
\item For $J=E_7$ 
\be\tilde{a}=-\frac{189}{32}(1+\epsilon)(458-915\epsilon+459\epsilon^2);\quad \tilde{c}=\tilde{a}+\frac{63}{16}(1+\epsilon).\ee 
\item For $J=E_8$ 
\be\tilde{a}=-\frac{45}{4}(1+\epsilon)(1304+3\epsilon(435\epsilon-869));\quad\tilde{c}=\tilde{a}+\frac{15}{2}(1+\epsilon).\ee
\end{itemize}
We can now notice that $\tilde{a}$ always has a minimum at $\epsilon=-\frac{1}{3}$. Analogously, the derivative of (\ref{higgs2}) vanishes at $\epsilon=-\frac{1}{3}$ so we conclude that the same will be true for the trial central charge of (the interacting sector of) the IR fixed point, which is just given by (\ref{higgs2}) minus $\tilde{a}$. The second derivative at $\epsilon=-\frac{1}{3}$ reads\footnote{Notice that this formula is valid for any $\CN=2$ SCFT with ADE global symmetry} (as before $C_i(J)$ denote the degree of the Casimir invariants) 
\be\sum_i\left(\frac{9}{4}C_i(J)^3-\frac{27}{16}C_i(J)^2\right)-\frac{9}{32}Dim(J)(3h(J)^2+2).\ee
One can check using the formulas reported in Section \ref{geomsec} that this quantity is always negative for $D_k^b(J)$ theories with $k>b$. We therefore conclude that the trial $a$ central charge is always maximized at $\epsilon=-\frac{1}{3}$ for $k>b$.

\section{Counting mass parameters}

The purpose of this section is to count mass parameters for $D_k^b(J)$ theories (besides the Casimirs of the $J$ global symmetry). These appear as complex structure deformations of dimension one. As a byproduct, we will relate this to the counting of mass parameters for $J^b(k)$ theories.  

Before proceeding with the case-by-case analysis, let us summarize our findings in the following table: 
\begin{table}[h]
\centering\begin{tabular}{|c|c|}
\hline
\text{Theory} & \text{Number of mass parameters} \\
\hline
$D_k^b(SU(N))$ & $GCD(b,k)-b+N-1$ \\
\hline 
$D_k(SO(2N))$ & $\frac{GCD(2N-2,k)+2}{2}$ for $\frac{2N-2}{GCD(2N-2,k)}$ odd; 1 for $k$ and $\frac{2N-2}{GCD(2N-2,k)}$ even; 0 for k odd \\
\hline 
$D_k^N(SO(2N))$ & $GCD(N,k)$ for $\frac{N}{GCD(N,k)}$ odd; 0 otherwise \\
\hline 
$D_k(E_6)$ & 6 for $k=0(mod12)$; 2 for $k=3,6,9(mod12)$; 0 for $k\neq0(mod3)$ \\
\hline 
$D_k^9(E_6)$ & 6 for $k=0(mod9)$; 0 otherwise \\
\hline 
$D_k^8(E_6)$ & 6 for $k=0(mod8)$; 2 for $k=4(mod8)$; 1 for $k\neq0(mod4)$ \\
\hline
$D_k^b(E_7)$ & 7 for $k=0(modb)$; 1 for k even and $k\neq0(modb)$; 0 for k odd \\
\hline 
$D_k^b(E_8)$ & 8 for $k=0(modb)$; 0 otherwise \\
\hline
\end{tabular} 
\caption{Number of mass parameters of $D_k^b(J)$ theories excluding the $J$ Casimirs.}
\label{massnum}
\end{table}

\noindent Notice in particular that for fixed $J$ the number of mass parameters just depends on $b$ and the value of $k$ modulo $b$.
The rank of the global symmetry group is always $r(J)$ plus the number written in the above table. 

\subsection{$D_k^b(SU(N))$ theories} 

When $b=N$ the allowed deformations are monomials of the form $u_{ij}x^it^j$ with $0<j<k$ and $i<N-1$. We are interested in counting terms such that $D(u_{ij})=1$. Since the dimension of $x$ is one, the problem is equivalent to determine the values of $j$ such that $t^j$ has integer dimension. This leads to the equation 
$$\frac{N}{k}j=n$$ 
for some positive integer $n$. Since $j<k$, there are solutions only when $GCD(N,k)\neq1$. The allowed values of $j$ are the multiples of $\tilde{k}=k/GCD(N,k)$ smaller than k and there are exactly $GCD(N,k)-1$ of them. 

The case $b=N-1$ is very similar. The set of allowed deformations is the same as before with the addition of the term $ut^k$. Since the dimension of $u$ is always one, we conclude that the global symmetry is always at least $SU(N)\times U(1)$ in this class of models. Other mass parameters are found again by imposing that $t^j$ (with $j<k$) has integer dimension: 
$$\frac{N-1}{k}j=n.$$
Following the same argument given before we find $GCD(N-1,k)-1$ solutions, for a total of $GCD(N-1,k)$ mass parameters. 
The final formula valid for all $D_k^b(SU(N))$ theories is then 
\be\label{masssun} GCD(b,k)-b+N-1\ee

\subsection{$D_k^b(SO(2N))$ theories}

In the case $b=2N-2$ the SW curve with all deformations turned on is 
$$x^{2N}+x^2t^k+\sum_{i,j}u_{ij}x^{2i}t^j+P(t)^2=0,$$ 
where $P(t)$ is a polynomial in $t$ of degree $\lfloor k/2\rfloor$ ($\lfloor\dots\rfloor$ denotes the integer part). We are interested only in terms with $i,j>0$. Notice that when $k$ is even, the polynomial $P(t)$ has exactly degree $k/2$ and consequently the curve includes the term $u^2t^k$. Since $x$ has dimension one, $u$ is always a mass parameter and consequently the global symmetry of the theory can be just $SO(2N)$ (with no enhancement) only for $k$ odd. Similarly to the $SU(N)$ case, other mass parameters can be identified by demanding that $D(t^j)$ is an odd integer. This imposes the constraint 
$$\frac{2N-2}{k}j=2n+1$$
for some nonnegative integer $n$. This equation has solutions only when $\tilde{N}=2N-2/GCD(2N-2,k)$ is odd, which in particular implies that $GCD(2N-2,k)$ (and consequently $k$ as well) is even. The number of solutions is equal to the number of integers $j<k$ of the form $(2m+1)\tilde{k}$ (where $\tilde{k}=k/GCD(2N-2,k)$) and there are precisely $GCD(2N-2,k)/2$ of them. 
In summary, when $k$ is odd the symmetry is exactly $SO(2N)$, when both $k$ and $\tilde{N}$ are even there is one mass parameter, when $\tilde{N}$ is odd there are $GCD(2N-2,k)/2+1$ mass parameters. 

In the case $b=N$ the SW curve reads 
$$x^{2N}+t^2k+\sum_{i,j}u_{ij}x^{2i}t^j+P_{k-1}(t)^2=0,$$
where $P_{k-1}(t)$ is a polynomial of degree $k-1$ in $t$. $i$ and $j$ are strictly positive and $j<2k$. First of all we notice that among the coefficients of $P_{k-1}(t)$ there are no mass parameters unless $k$ is a multiple of $N$ (if this is the case, there is exactly one parameter of dimension one): this follows from the fact that all the terms appearing in $P_{k-1}(t)$ have dimension $N$ and the dimension of $t$ is $N/k$. In order to count $u_{ij}$'s of dimension one we should impose the constraint 
$$\frac{N}{k}j=2n-1$$
for some nonnegative integer $n<N$. This equation of course implies that $\tilde{N}=N/GCD(N,k)$ is odd. All in all, we find a total of $GCD(N,k)$ mass parameters whenever $\tilde{N}$ is odd and zero otherwise. 

In conclusion, for $J=SO(2N)$ we find that whenever $b/GCD(b,k)$ is odd the number of mass parameters can be expressed in terms of $GCD(b,k)$. If this constraint is not satisfied, we find a single mass parameter for $b=2N-2$ and $k$ even and zero otherwise. 

\subsection{$D_k^b(E_6)$ theories}

In the exceptional case it is more convenient to use the Type IIB geometry. For $J=E_6$ and $b=12$ the Calabi-Yau geometry is 
$$x_1^2+x_2^3+x_3^4+t^k=0$$
and the allowed deformations are 
$$u_{ijl}x_2^ix_3^jt^l\quad i=0,1;\; j=0,1,2;\; l<k$$ 
For $l=0$ these represent the Casimir invariants of the $E_6$ global symmetry and have dimension 2,5,6,8,9 and 12 respectively. We then immediately conclude that the dimension of $u_{ijl}t^l$ is always equal to the degree of one of the $E_6$ Casimirs. Since $u_{ijl}$ (with $l>0$) is a mass parameter if and only if it has dimension one, we conclude that the dimension of $t^l$ has to be equal to one of the exponents of $E_6$ (degree of the Casimirs minus one). We should therefore find all integers $l$ such that 
$$\frac{12}{k}l=1,4,5,7,8,11.$$
Clearly there are 6 solutions when $k=0(mod12)$, 2 when $k=0(mod3)$ but $k\neq0(mod12)$ and zero otherwise. 

For $b=9$ the allowed deformations are 
$$u_{ijl}x_2^ix_3^jt^l\quad i=0,1;\; j=0,1,2,3;\; l<k.$$ 
The parameter $u$ is a CB operator of dimension three and clearly does not contribute to the counting. The new ingredient with respect to the previous case is the presence of the terms $u_{03l}x_3^3t^l$. The dimension of $u_{03l}$ is $$D(u_{03l})=3-\frac{9l}{k}$$ 
and there is exactly one value of $l$ such that $u_{03l}$ is a mass parameter if and only if $k=0(mod9)$. The analysis for the other terms is analogous to the $b=12$ case and we have the equation 
$$\frac{9}{k}l=1,4,5,7,8,11$$ 
and clearly there are no solutions if $k\neq0(mod9)$. If $k$ is a multiple of 9 we find five solutions. In conclusion we find 6 mass parameters when $k=0(mod9)$ and zero otherwise. 

Finally, for $b=8$ we have the deformations 
$$u_{ijl}x_2^ix_3^jt^l\quad i=0,1,2;\; j=0,1,2;\; l<k.$$ 
The case $i<2$ is analogous to those discussed previously and leads to the equation 
$$\frac{8}{k}l=1,4,5,7,8,11$$
which has four solutions when $k=0(mod8)$, one when $k=4(mod8)$ and zero otherwise. 
The remaining cases are $u_{20l}x_2^2t^l$ and $u_{21l}x_2^2x_3t^l$. $u_{210}$ is a mass parameter for any $k$ and $u_{20l}$ has dimension one for 
$$l=\frac{3k}{8}$$ 
which is clearly an integer only when $k=0(mod8)$. We conclude that there are 6 mass parameters for $k=0(mod8)$, 2 when $k=4(mod8)$ and 1 otherwise. 

\subsection{$D_k^b(E_7)$ theories} 

The case $b=18$ is very similar to the case $b=12$ of the previous section: we should find all integers $l<k$ such that the dimension of $t^l$ is an exponent of $E_7$.This leads to the equation 
$$\frac{18}{k}l=1,5,7,9,11,13,17$$
which has 7 solutions when $k=0(mod18)$, 1 solution when $k$ is even but is not a multiple of 18 and zero otherwise. 

For $b=14$ the allowed deformations are of the form
$$u_{ijl}x_2^ix_3^jt^l\quad i=0,1,2;\; j=0,1,2;\; l<k.$$ 
The parameters $u_{12l}$ have dimension 
$$D(u_{12l})=4-\frac{14}{k}l$$ 
and provide exactly one mass parameter when $k$ is a multiple of 14. All other $u_{ijl}$ parameters are present also in the case $b=18$ and according to the usual argument correspond to mass parameters if 
$$\frac{14}{k}l=1,5,7,9,11,13,17.$$ 
There are no solutions for $k$ odd, one solution if $k$ is even but is not a multiple of 14 and 6 for $k=0(mod14)$. Overall we find 7 mass parameters when $k$ is a multiple of 14, 1 when it is even but $k\neq0(mod14)$ and zero otherwise. 

\subsection{$D_k^b(E_8)$ theories} 

Finally, let's consider the $J=E_8$ case. For $b=30$ we have the equation 
$$\frac{30}{k}l=1,7,11,13,17,19,23,29$$ 
which has 8 solutions when $k=0(mod30)$ and zero otherwise. 

In the case $b=24$ we have the deformations $u_lx_3^4t^l$, which provide one mass parameter when $k=0(mod24)$, besides those which are there also in the $b=30$ case: 
$$u_{ijl}x_2^ix_3^jt^l\quad i=0,1;\; j=0,1,2,3;\; l<k.$$ 
These correspond to mass parameters if the following condition is satisfied: 
$$\frac{24}{k}l=1,7,11,13,17,19,23,29.$$ 
There are 7 solutions when $k=0(mod24)$ and zero otherwise. In total we find 8 mass parameters when $k$ is a multiple of 24. 

For $b=20$ we have again the deformations 
$$u_{ijl}x_2^ix_3^jt^l\quad i=0,1;\; j=0,1,2,3;\; l<k$$
which lead to mass parameters when 
$$\frac{20}{k}l=1,7,11,13,17,19,23,29.$$ 
There are 6 solutions when $k$ is a multiple of 20. We also find the terms 
$$u_{jl}x_2^2x_3^jz^l\quad j=0,1;\; l<k$$ 
and the equations determining the existence of mass parameters is 
$$\frac{20}{k}l=3,9$$
which leads to two extra solutions, for a total of 8, when $k$ is a multiple of 20. There are no solutions if $k$ is not a multiple of 20. 

\subsection{Counting mass parameters for $J^b(k)$ theories}

As we have seen, for $D_k^b(J)$ theories mass parameters (besides the Casimirs of $J$) arise in the geometric setup as deformation terms of the form $u_{i,j,l}x_2^ix_3^jt^l$ with $l>0$. As we have noticed in Section (\ref{scftj}), the scaling dimensions of the coordinates $x_i,z$ for $J^b(k)$ theories can be obtained by multiplying (\ref{scaledim}),(\ref{scalet}) by $\frac{k}{k+b}$. We therefore conclude that the parameter $u_{i,j,l}$ appearing in the deformation term $u_{i,j,l}x_2^ix_3^jz^l$ has dimension $\frac{k}{k+b}$, whenever the corresponding parameter for $D_k^b(J)$ theories has dimension one. Since the dimension of $z$ is always $\frac{b}{k+b}$, we immediately find that $u_{i,j,l-1}$  for $J^b(k)$ theory has dimension exactly one and hence is a mass parameter. At this stage a useful observation is the following: if $x_2^ix_3^jt^l$ is not set to zero by the relations generating the ideal $I_W$ for $D_k^b(J)$ theories (which include the relation $t\partial W/\partial t=0$), then $x_2^ix_3^jz^{l-1}$ will not be in the ideal for $J^b(k)$ theories (which is generated by the relation $\partial W/\partial z=0$; the other relations are identical to those of $D_k^b(J)$ with $t$ replaced by $z$). This allows us to conclude that $u_{i,j,l}$, with $l>0$, is a mass parameter for $D_k^b(J)$ theory if and only if $u_{i,j,l-1}$ is a mass parameter for $J^b(k)$. Using this observation it is now easy to count mass parameters for $J^b(k)$ theories: their number is equal to that appearing in Table (\ref{massnum}) for $D_k^b(J)$ theories. In particular, we can notice that the constraint on $k$ so that $J^b(k)$ theories have no mass parameters (given in \cite{Xie:2016evu}) is equal to the corresponding constraint appearing in Table (\ref{massnum}), i.e. the requirement that the global symmetry is exactly $J$ with no enhancement.

\bibliographystyle{ytphys}

\end{document}